\documentclass[journal=jacsat,manuscript=article]{achemso}
\usepackage{color}

\usepackage[version=3]{mhchem} 

\author{Yuan-Fei Gao}
\affiliation{State Key Laboratory of Superlattices and Microstructures, Institute of Semiconductors, \& Center of Materials Science and Optoelectronics Engineering, University of Chinese Academy of Sciences, Chinese Academy of Sciences, Beijing 100083, People’s Republic of China}
\alsoaffiliation{Beijing Academy of Quantum Information Sciences, Beijing 100193, People’s Republic of China}
\author{Jia-Min Lai}
\author{Zhen-Yao Li}
\author{Ping-Heng Tan}
\affiliation{State Key Laboratory of Superlattices and Microstructures, Institute of Semiconductors, \& Center of Materials Science and Optoelectronics Engineering, University of Chinese Academy of Sciences, Chinese Academy of Sciences, Beijing 100083, People’s Republic of China}
\author{Chong-Xin Shan}
\email{cxshan@zzu.edu.cn}
\affiliation{Henan Key Laboratory of Diamond Optoelectronic Materials and Devices, Key Laboratory of Material Physics, Ministry of Education, School of Physics and Microelectronics, Zhengzhou University, Zhengzhou 450052, People’s Republic of China}
\author{Jun Zhang}
\email{zhangjwill@semi.ac.cn}
\affiliation{State Key Laboratory of Superlattices and Microstructures, Institute of Semiconductors, \& Center of Materials Science and Optoelectronics Engineering, University of Chinese Academy of Sciences, Chinese Academy of Sciences, Beijing 100083, People’s Republic of China}
\title [An \textsf{achemso} demo] {Local laser heating effects in diamond probed by photoluminescence of $\mathrm{\mathbf{SiV^-}}$ centers at low temperature}

\abbreviations{IR,NMR,UV}
\keywords{American Chemical Society, \LaTeX}

\begin{document}

\begin{tocentry}
\includegraphics[width=\linewidth]{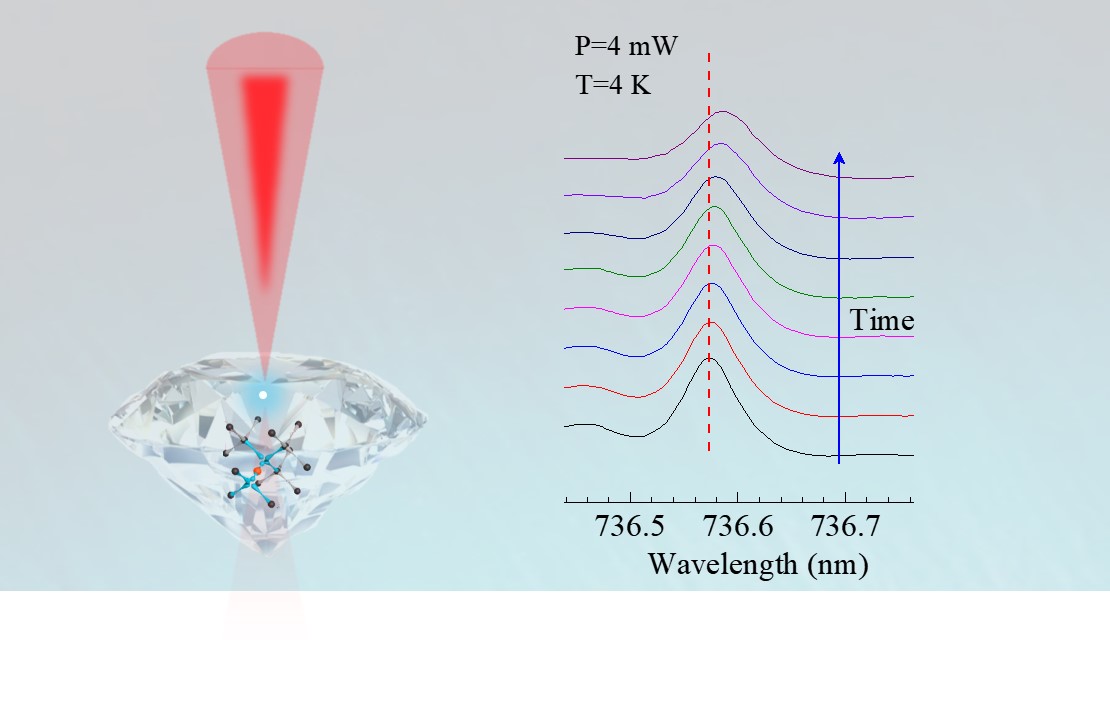}




\end{tocentry}

\begin{abstract}
Diamond is generally considered to have high thermal conductivity, 
 so little attention has been paid to the laser heating effects at low excitation power.
 However, defects during the growth process can result in a great degradation of thermal conductivity, especially at low temperatures.
 Here, we observed the dynamic redshift and broadening of zero phonon line (ZPL) of silicon-vacancy ($\mathrm{SiV^-}$) centers in diamond in the experiment.
 Utilizing the intrinsic temperature response of the fine structure spectra of $\mathrm{SiV^-}$ as a probe,
 we confirmed that the laser heating effect appears and the temperature rising results from high defect concentration.
 By simulating the thermal diffusion process, we have estimated the thermal conductivity of around 1 $W/(m \cdot K)$ at the local site, which is a two-order magnitude lower than that of single-crystal diamond. 
 Our results provide a feasible scheme for characterizing the laser heating effect of diamond at low temperatures.
\end{abstract}

\section{Introduction}
Diamond is generally considered to possess high thermal conductivity due to low mass of carbon atoms, strong interatomic bonding in lattice, and low anharmonicity of the interatomic potential\cite{1-RN9941}.
It is considered to be a promising material used for heat dissipation of high-power semiconductor devices which are key to delivering high-efficiency energy conversion in power electronics systems\cite{2-RN9967}. 
Therefore, little attention has been paid to the laser heating effect of diamond.
However, its thermal conductivity strongly depends on impurity concentration and lattice imperfections. 
Nitrogen could present in a large concentration in diamonds, which reduces the thermal conductivity by several times\cite{3-RN9942}.
It is well known that heating due to the decrease of thermal conductivity will degrade the performance of semiconductor devices\cite{2-RN9967}.
In addition, diamond shows strong isotope effect: diamond grown with the isotopically enriched $C^{12}$ could increase the value of thermal conductivity up to $50\%$ at room temperature\cite{4-RN9969,5-RN9968}.
Thus, the large uncertainty of thermal conductivity hinders the application of diamonds in optoelectronic devices, especially in extreme environments.
Recently, thermal conductivity of three synthetic single-crystal diamonds was measured with high accuracy at temperatures from 6 to 410 K \cite{6-RN9948}. 
However, the thermal conductivity was measured by a steady-state longitudinal heat flow method.
This measurement method is not suitable for thermal conductivity measurement locally at low temperatures, mainly due to the inaccuracy of temperature difference measurement.
Therefore, it is important to find an alternative temperature measurement method for estimating thermal conductivity at low temperatures.

Nanoscale sensing based on atom-like solid-state systems is promising in metrology \cite{7-RN9949,8-RN9950,9-RN9951}.
Diamond has lots of advantages in local temperature sensing.
It is chemically and physically inert and has low toxicity which is considered  to have a minimal effect on the cell functionality\cite{10-RN9898}.
There are many color centers with excellent optical properties in diamond, such as nitrogen-vacancy (NV) center\cite{12-RN6403,11-RN8010} and silicon-vacancy (SiV) center\cite{13-RN8651,14-RN5152}.
The color centers have attracted particular interest for their robustness to extreme environments and for their ability to be localized within nanometers. 
 The electron spin of NV centers could be used to probe magnetic, temperature, electric, and strain fields within nanometers and high sensitivity\cite{7-RN9949,8-RN9950,9-RN9951,15-RN4908,16-RN9953}.
While an NV center-based sensor provides outstanding performance, it has a limitation since it requires the application of microwave radiation.
Microwave radiation cannot be focused below a small size and can produce heating effects.
A possible alternative to the NV center may be silicon-vacancy ($\mathrm{SiV^-}$) center or the recently discovered germanium vacancy ($\mathrm{GeV^-}$) center \cite{17-RN6280,18-RN6043}, which has optical spectra dominated by a zero-phonon line( ZPL), thus, offering a possibility to measure temperature by optical method.
At low temperatures, the $\mathrm{SiV^-}$ center’s ZPL has been observed to split into a four-line fine structure, which is regarded as a “spectral fingerprint” of the SiV center in a high-quality, low-stress diamond.
The origin of the fine structure can be attributed to a level scheme where both excited and ground states are split into two substates\cite{19-RN6153,20-RN7795,21-RN4053}. 
The fluorescence of $\mathrm{SiV^-}$ centers is predominantly concentrated in the purely electronic transition, and the ZPL features a room temperature width of down to 0.7 nm\cite{20-RN7795}.
Extremely narrow linewidth provides a high sensitivity for temperature measurement which is not available with other fluorescent probes. 
Thus, we can measure the temperature evolution using micro-PL spectra of $\mathrm{SiV^-}$ center as a probe at low temperature.

Here, by studying the fine structure spectrum of $\mathrm{SiV^-}$ centers at low temperatures, we observe the PL spectra are dependent on the excitation power. 
Under high-power excitation, the spectra show obvious redshift and broadening.
By calibrating with the intrinsic temperature dependence of the $\mathrm{SiV^-}$  spectra, we confirm that the evolution of the spectra resulted from laser heating effects in the center of the laser spot. 
Through theoretical analysis of the experiment, the temperature rising in the center of the spot is determined by the relatively low thermal conductivity.
By measuring the defect concentration distribution, we confirmed the large doping of defects and the lattice imperfection in the initial growth greatly reduce the thermal conductivity.
By simulating the thermal diffusion process, we estimate the thermal conductivity of diamond has reduced by nearly two orders of magnitude than that of high-quality single-crystal diamond. 
It is shown that the fine structure spectrum of $\mathrm{SiV^-}$  centers could be a sensitive probe for laser heating effects in diamonds. 
In addition, it is an alternative way to estimate the thermal conductivity of diamonds at low temperatures. 

\section{Results and discussion}
\begin{figure}
    \centering
    \includegraphics[width=0.8\linewidth]{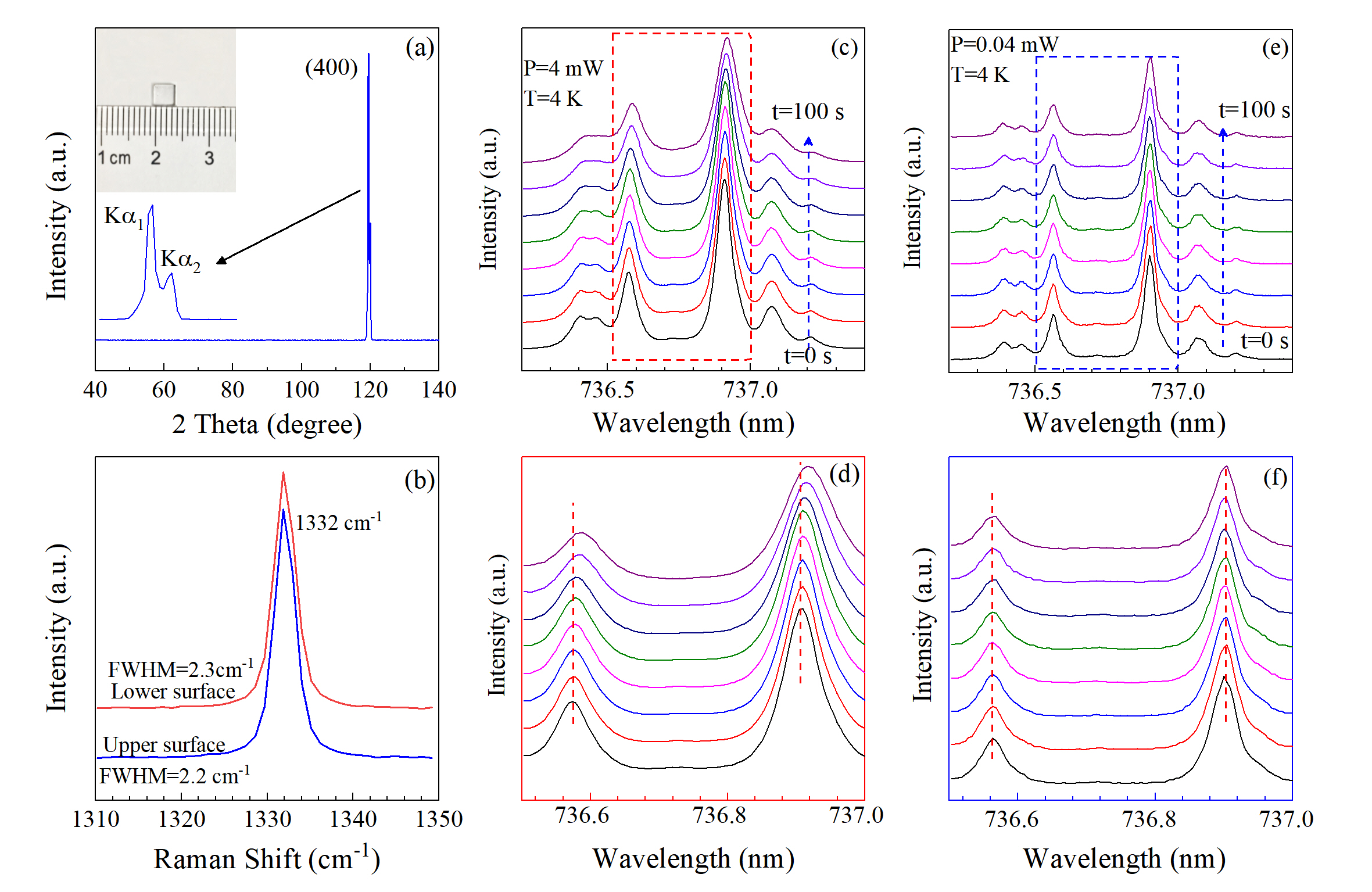}
    \caption{The basic characterization of the CVD diamond and the time evolution of the fine structure spectra of $\mathrm{SiV^-}$  centers in diamond at 4 K under different excitation powers. (a)~X-ray diffraction (XRD) pattern; the inset shows the optical image. (b)~Raman spectra at the lower and upper surface. The lower and upper surfaces of the sample correspond to the beginning and end stages of sample growth, respectively.(c)~and (f)~Excitation powers are 4 mW and 0.04 mW at the lower surface, respectively. The spot area is about 4 $\mu m^2$. (d)~and (f)~are enlarged for clarity.}
\end{figure}
Figures 1(a) and 1(b) show the basic characterization of the CVD diamond.
The XRD shows only one peak located at about 119.9 $\mathrm{cm^{-1}}$ , which proves that the diamond is mono-crystalline.
The $(400)$ peak consists of two sub-peaks, which comes from the Cu $K\alpha_{1}$ and Cu $K\alpha_{2}$ line;
The splitting of the XRD peak suggests high crystal quality of the CVD diamond\cite{22-RN4418}.
The inset shows the optical image in which the sample is nearly colorless and transparent.The absorption spectrum is shown in Fig.S1.There is a sharp absorption edge at around 5.5 eV, which corresponds to the band-gap of diamond.
Fig.1(b) shows the Raman spectra of the CVD diamond located at 1332 $\mathrm{cm^{-1}}$ with a full width at half maximum (FWHM) of 2.3 $\mathrm{cm^{-1}}$ at the lower surface and 2.2 $\mathrm{cm^{-1}}$ at upper surface, respectively.
Therefore, it suggests that the sample grown by MPCVD has a high crystalline quality.
In homoepitaxial CVD diamond films with high crystalline quality, the ZPL of the $\mathrm{SiV^-}$ centers splits into four fine structure peaks at liquid helium temperatures.
Figs. 1(c) and 1(e) display the evolution of ZPL of the $\mathrm{SiV^-}$  ensemble under different excitation power at 4 K.
Here, we focus on the spectral properties at liquid-helium temperature.
We note that the evolution of the fine structure spectra under 4 mW excitation has a significant redshift and broadening. Simultaneously, the PL intensity  decreases.
Under lower excitation power (0.04 mW), fluorescence spectra hardly changes over time.
Therefore, we suspect that there is a laser heating effect in the high-power excitation process, which leads to a significant change in the fine structure spectra. 
\begin{figure}
    \centering
    \includegraphics[width=0.8\linewidth]{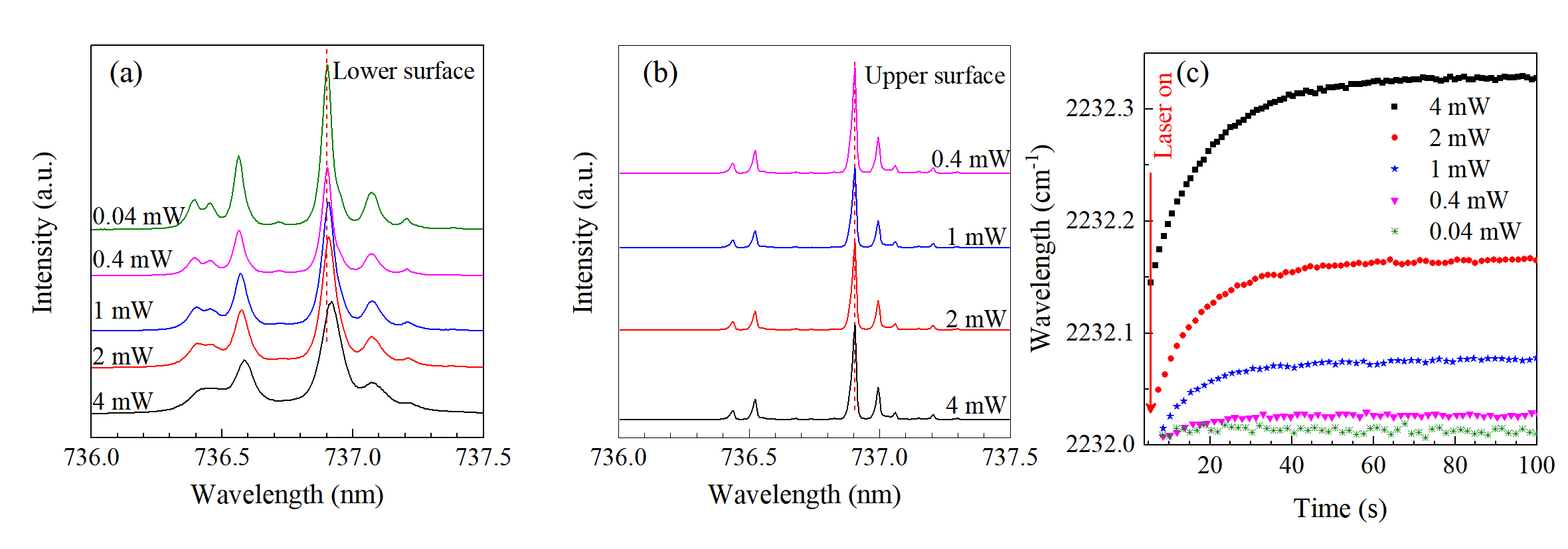}
    \caption{The comparison of the fine structure spectra on the upper and lower surfaces. (a)~and (b)~Stable PL spectra of $\mathrm{SiV^-}$  centers in diamond under different excitation powers on the lower and upper surfaces, respectively.
    (c)~the redshift of the strongest peak of the fine structure spectrum in the lower surface with time. The vertical axis is expressed as a relative wavenumber ($\mathrm{cm^{-1}}$) with an excitation wavelength of 632.8 nm.}
\end{figure}

The heating effect goes against the perception that diamond possess high thermal conductivity. 
Then, we made a detailed measurements on the sample.
Figs. 2(a) and 2(b) show the fine structure spectra of $\mathrm{SiV^-}$  centers on the lower and upper surfaces at 4 K under different excitation power, respectively. 
Since the spectra evolve over time, the spectra shown here in the lower surface are the result after the final thermal stabilization. 
The fine structure spectra on the upper surface have no dependence on the power, while the spectra on the lower surface show a significant power dependence. 
The fine structure spectra in the upper and lower surfaces are significantly different and the linewidth of spectra in the lower surface is much wider than that in the upper surface.
A crucial prerequisite to observe the $\mathrm{SiV^-}$  fine structure is a low strain crystal environment and defect concentration \cite{23-RN4878,24-RN6127}. 
When the strain is large or the defect concentration is relatively high, the fine structure would widen and disappear. 
Therefore, the laser heating effect in the lower surface may be related to the residual stress and defect concentration \cite{20-RN7795}.
To study the laser heating effect, the evolution of the fine structure spectra under different excitation powers were measured. 
Fig.~2(c) shows the evolution of the peak position of the strongest peak over time.
The peak position undergoes a significant redshift when the exciting power is greater than 0.04 mW. 
Under different excitation powers, the evolution time of the spectra is almost on the order of tens of seconds, which means that the thermal diffusion process is very slow.
\begin{figure}
    \centering
    \includegraphics[width=0.8\linewidth]{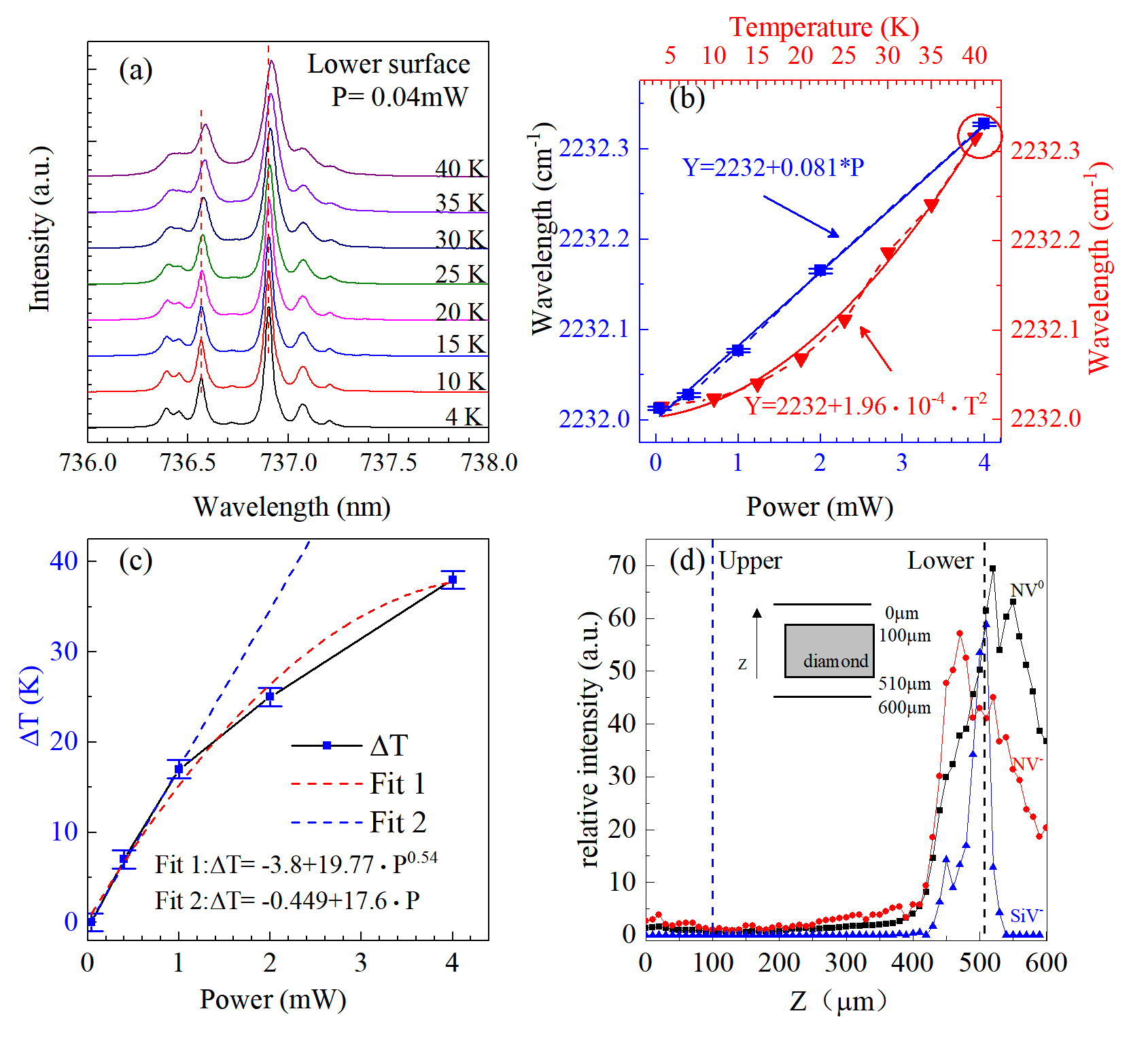}
    \caption{(a)~The intrinsic temperature-dependent PL spectra of $\mathrm{SiV^-}$ in the lower surface under 0.04 mW excitation.
    (b)~The strongest peak position under different excitation power (blue) at 4 K, temperature dependence of the peak position of the strongest peak under 0.04 mW excitation (red). 
    (c)~Temperature changes at the center of the laser spot under different excitation power.
    (d)~Variation of defect concentration along the growth direction of the sample normalized by Raman peak at 1332 $\mathrm{cm^{-1}}$.}
    \label{fig:my_label}
\end{figure}

In semiconductors, the heating of the lattice strongly influences the luminescence spectra. 
This is manifested in two ways in the PL spectra.
Firstly, the spectral distribution of PL intensity changes due to the increase in kinetic energy of photo-excited carriers\cite{25-RN9899}.
Secondly, the peak position of photoluminescence shows appreciable shift with temperature, which result from the variation in the energy of electronic states produced by anharmonic interaction of electrons and phonons\cite{26-RN6332}.
The temperature dependence of the energy gap of semiconductors is usually described by a semiempirical relation known as Varshni’s equation\cite{27-RN9957}.
In Fig.~3(a), the intrinsic temperature dependence of the $\mathrm{SiV^-}$ centers were measured under lower excitation power.
We found the evolution of spectra with temperature is similar to that with time in Fig. 1(c).
Fig. 3(b) shows the peak shift (blue) of the strongest peak under different excitation powers and the intrinsic peak shift (red) with temperature under low-power excitation.
The stable peak position under 4 mw excitation is almost consistent with the intrinsic peak position at 40 K, thus the temperature at the center of the laser spot under 4 mW excitation is about 40 K and the laser spot center does have a noticeable temperature rise.
According to the work of Bergman et al.\cite{28-RN7944}, the PL redshift is attributed to laser heating and heat trapping.
When the laser beam is focused on the surface of diamond, part of the optical energy is absorbed by defects, causing transition processes. 
Only the energy in the nonradiative transition process is transformed into heat energy.
According to the work of Yang et al. \cite{29-RN3582}, under an excitation power of $P$,the power density of heat $P_h$ can be expressed as
\begin{equation}
    P_{h}=L(1-\eta)FP 
\end{equation}

Here, $\eta$ is the total luminescence quantum yield, $F$ is the fraction of incident light absorbed by defects and  $L$ is a proportionality constant. 
When the laser power is steady, there is a relation between the power density of heat ($P_h$ ) and temperature at the heating area ($T$):
\begin{equation}
    P_{h}\approx \kappa(T-T_0)
\end{equation}
where $\kappa$ is the coefficient, which is directly proportional to the thermal conductivity, and $T_0$  is the sample holder temperature. Thus, it has
\begin{equation}
    L(1-\eta )FP\approx\kappa (T-T_0)\Rightarrow T\approx \mu P+T_0
\end{equation}
Where $\mu$ is a coefficient, and $\mu =L(1-\eta )F/\kappa$.
Under the same laser power, a larger $\mu$ leads to a stronger local heating effect.
In Fig. 3(c), we found the local heating effect does not increase linearly with excitation power. 
We can fit the data with a sublinear function (Fit 2: $\Delta T=-3.8+19.77\cdot P^{0.54}$ ) very well.
The main reason is that: with the high-power excitation, the thermal conductivity increases with temperature\cite{6-RN9948}.
At room temperature, the thermal conductivity of diamond is much higher than that at 4 K and the larger thermal conductivity leads to the smaller $\mu$.
To verify the theory is correct, we performed the same experiments at room temperature, and did not observe the redshift and broadening whether on the upper or lower surface of diamond.
Previously, the temperature dependence of the interband transition energies can be described by a Bose-Einstein-type expression\cite{30-RN9900}.
The reason of peak shift and the exact temperature dependence have been discussed in the literature\cite{27-RN9957,31-RN7709}. Lattice expansion and temperature-dependent electron lattice interaction led to a temperature-dependent bandgap in diamond.
However, in Neu’s work\cite{32-RN8547}, they discuss the temperature dependence of fine structure in more detail and make corrections to previous work.
They found their measured data are consistent with a  temperature dependence. Our measurement results and fitting are consistent with them, as presented in Fig. 3(b).
Under the same conditions, local heating effect were only observed on the lower surface.
In Eq 3, $\mu$ determines the magnitude of the laser heating effect under the same excitation power, and it is determined by the absorption coefficient and thermal conductivity. 
The larger the absorption coefficient and the smaller the thermal conductivity, the larger will $\mu$ be.
Therefore, the reason for the different phenomena observed on the upper and lower surfaces comes from the large difference in thermal conductivity.
Fig. 3(d) shows the distribution of defect concentration in the growth direction of the sample. 
The defect concentration is estimated by PL intensity.
We found the defect concentration on the lower surface is significantly higher than that on the upper surface. 
The thermal conductivity of diamond depends strongly on impurity concentration and crystal lattice imperfection\cite{6-RN9948}.
For example, nitrogen, which can present in a large concentration in diamonds, up to 0.25 reduces the thermal conductivity by several times\cite{33-RN9959}.
High defect concentration absorbs more laser and turns into heat and low thermal conductivity leads to greater temperature rise.
Therefore, the root cause is the high defect concentration on the lower surface.
\begin{figure}
    \centering
    \includegraphics[width=0.8\linewidth]{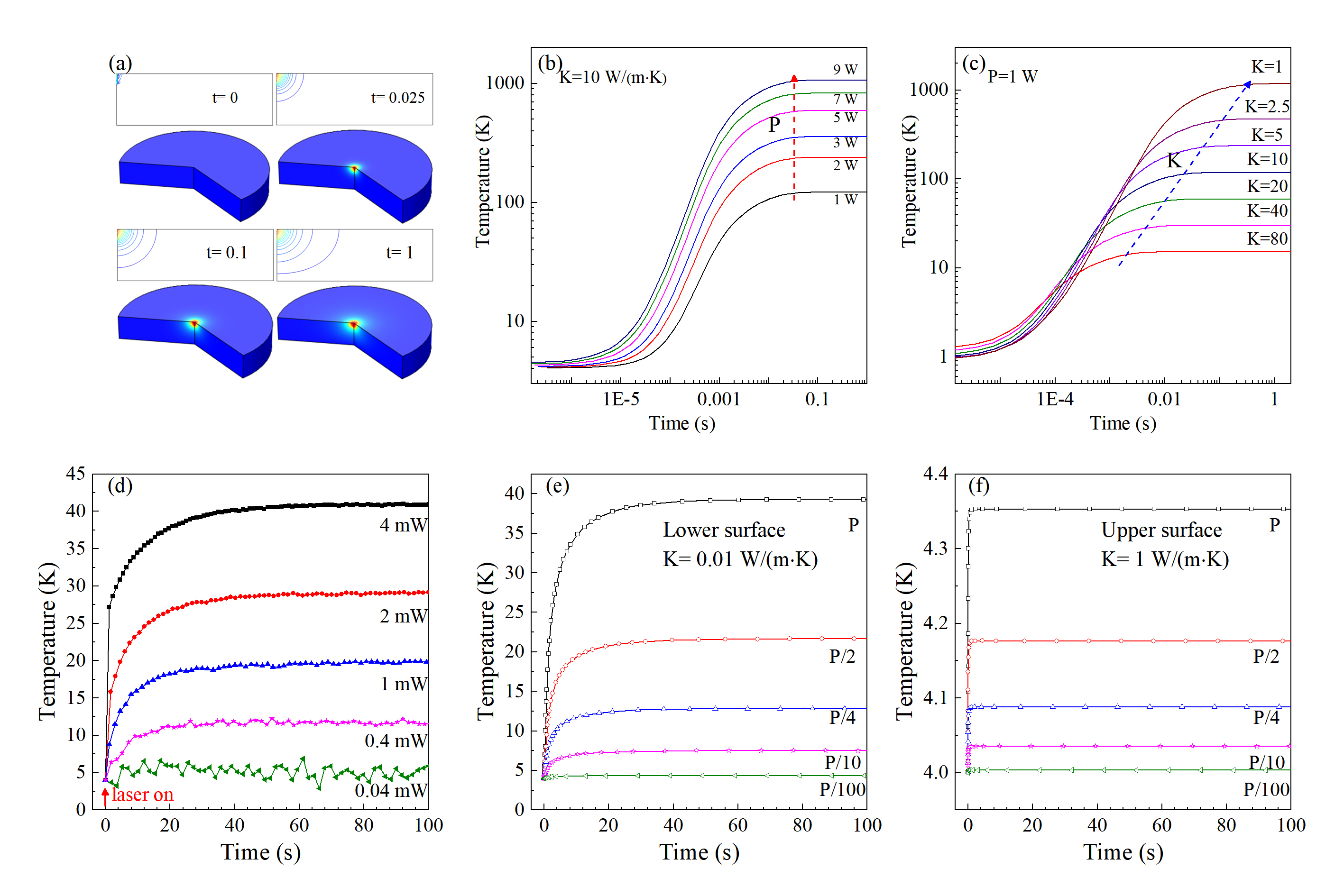}
    \caption{The diffusion process simulation for laser heating. 
    (a) Isotherm distribution in cross-section and three-dimensional temperature distribution at different times, excitation power, and thermal conductivity is 1 W and 10 $W/(m \cdot K)$.
    (b) The evolution process of laser spot center temperature under different excitation powers when the thermal conductivity is 10 $W/(m \cdot K)$.
    (c) The evolution process of spot center temperature at different thermal conductivity when the excitation power is 1 W.
    The simulation of the relationship between thermal conductivity and equilibrium time. 
    (d) the evolution of the temperature of the spot center under different excitation power calibrated by the temperature-dependent photoluminescence spectra.
    (e) and (f) simulate the temperature evolution process of the spot center under different thermal conductivity, respectively.
    Among them, $1~W/(m\cdot K)$ and $100~W/(m\cdot K)$ are applicable to the lower surface and the upper surface of the diamond sample.}
\end{figure}

In order to have a more intuitive understanding of the laser heating process, we used a simplified model to simulate the thermal diffusion process shown in Fig.4.The detailed simulation process is shown in Fig.S2 of the supplementary materials.
Since the measurement process cannot quantify the absorbed energy, the simulation process here is qualitatively judged. 
The spot area of the laser is much smaller than the sample size, so the two-dimensional axisymmetric structure is used to simplify the thermal diffusion process. 
Fig. 4 (a) shows the isotherm distribution of the cross-section during the temperature rise.
Temperature spreads around the center of the spot and quickly reaches stability.
Next, we simulated the effects of excitation power and thermal conductivity on the thermal diffusion process as shown in Fig. 4(b) and 4(c).
Consistent with Eq.3 above, regardless of the change in thermal conductivity, the steady temperature varies nearly linearly with the excitation power. 
Also, the equilibrium time of the thermal diffusion process does not change significantly under different excitation power which is consistent with the experimental results. 
The thermal diffusion process was observed to be a relatively slow process during the experiment.
The relationship between thermal conductivity and thermal equilibrium time is simulated.
Fig. 4(d) shows the evolution of temperature at the center of the laser spot with different excitation powers over time, which is calibrated by the temperature-dependent photoluminescence in Fig. 3(b). 
The equilibrium time of the thermal diffusion process is strongly dependent on the thermal conductivity of the sample.
Here, assuming that the diamond on the upper surface is an ideal high-quality single crystal diamond, and its thermal conductivity is set to 100 $W/(m\cdot K)$.
Through simulation, it can be found that there is almost no temperature change on the upper surface in Fig. 4(f).
For the lower surface of the diamond, when the thermal conductivity is set to 1 $W/(m\cdot K)$, the temperature evolution in the center of the spot is almost the same as the experiments in Fig. 4(e).
Therefore, we confirm that the high defect concentration on the lower surface results in the thermal conductivity being greatly reduced to nearly two orders of magnitude.

By studying the fine structure spectrum of $\mathrm{SiV^-}$ centers in diamonds at low temperatures, we observed the dependence of the spectra on the excitation power.
Under high-power excitation, the spectra show obvious redshift and broadening.
By studying the temperature dependence of the spectra under low-power excitation, we confirm that it is related to laser heating effects at the center of the laser spot.
Through theoretical analysis, the temperature rising in the center of the spot is determined by the low thermal conductivity of the diamond at low temperature. 
The large introduction of defects and the lattice imperfection in the initial growth greatly reduce the thermal conductivity of diamond on the lower surface. 
It is shown the fine structure spectrum of $\mathrm{SiV^-}$ centers could be a sensitive probe for temperature measurement at low temperature due to extremely narrow line width. 
Through simulation, we estimate that high defect conductivity concentrations degrade thermal conductivity by nearly two orders of magnitude. 
This method could be used to characterize the laser heating effects, which can be used to estimate the thermal conductivity of diamond.

\subsection{Experiments}
The diamond wafer was homoepitaxially grown on a high-pressure and high-temperature Ib diamond substrate by microwave plasma-enhanced chemical vapor deposition. 
The detailed growth process has been depicted elsewhere \cite{34-RN9965}. Briefly, the precursors used for growth of the diamond are $\mathrm{CH_4}$ and $\mathrm{H_2}$. 
The pressure in the growth chamber is nearly 330 mbar, and the substrate temperature is nearly 950 °C.
After growth, the CVD diamond wafer was removed from the substrate by laser cutting and then used for optical measurements.The length, width, and height of the sample are 5 $mm$, 5 $mm$, and 0.41 $mm$, respectively.
The sample is cleaned with ethanol and acetone solutions before measurement to remove surface residues, respectively.
The doping of $\mathrm{SiV^-}$ centers is mainly related to the etching of hydrogen plasma during the growth process and the residual silicon contamination of the growth chamber.
Quantifying the silicon concentration is challenging as both the degree of contamination and the efficiency of the doping process are unknown. 
In addition, fine-structure spectra cannot be observed in high concentration samples at low temperature, thus the defect concentration is relatively low.Compared with a sample with a fixed injection concentration purchased from Element Six, we estimated that the concentration in our sample is about 100 ppm in the lower surface.
PL measurements on diamond samples were undertaken in backscattering geometry with a Jobin-Yvon HR800 system equipped with a liquid-nitrogen-cooled charge-coupled detector.
The ASPL measurements were undertaken with a 50 $\times$ long-working-distance objective lens (NA=0.5) and 
2400 lines $\mathrm{mm^{-1}}$ grating were used at low temperature.


\begin{acknowledgement}
We acknowledge support from Beijing Natural Science Foundation (JQ18014), National Basic Research Program of China (grant no.2016YFA0301200, 2017YFA0303401), Strategic Priority Research Program of Chinese Academy of Sciences (grant no. XDB28000000), and NSFC (51527901, U1804155, U1604263).
\end{acknowledgement}




\bibliography{acs-achemso}

\providecommand{\latin}[1]{#1}
\makeatletter
\providecommand{\doi}
  {\begingroup\let\do\@makeother\dospecials
  \catcode`\{=1 \catcode`\}=2 \doi@aux}
\providecommand{\doi@aux}[1]{\endgroup\texttt{#1}}
\makeatother
\providecommand*\mcitethebibliography{\thebibliography}
\csname @ifundefined\endcsname{endmcitethebibliography}
  {\let\endmcitethebibliography\endthebibliography}{}
\begin{mcitethebibliography}{35}
\providecommand*\natexlab[1]{#1}
\providecommand*\mciteSetBstSublistMode[1]{}
\providecommand*\mciteSetBstMaxWidthForm[2]{}
\providecommand*\mciteBstWouldAddEndPuncttrue
  {\def\EndOfBibitem{\unskip.}}
\providecommand*\mciteBstWouldAddEndPunctfalse
  {\let\EndOfBibitem\relax}
\providecommand*\mciteSetBstMidEndSepPunct[3]{}
\providecommand*\mciteSetBstSublistLabelBeginEnd[3]{}
\providecommand*\EndOfBibitem{}
\mciteSetBstSublistMode{f}
\mciteSetBstMaxWidthForm{subitem}{(\alph{mcitesubitemcount})}
\mciteSetBstSublistLabelBeginEnd
  {\mcitemaxwidthsubitemform\space}
  {\relax}
  {\relax}

\bibitem[Slack(1973)]{1-RN9941}
Slack,~G.~A. Nonmetallic crystals with high thermal conductivity. \emph{Journal
  of Physics and Chemistry of Solids} \textbf{1973}, \emph{34}, 321--335\relax
\mciteBstWouldAddEndPuncttrue
\mciteSetBstMidEndSepPunct{\mcitedefaultmidpunct}
{\mcitedefaultendpunct}{\mcitedefaultseppunct}\relax
\EndOfBibitem
\bibitem[Zhang \latin{et~al.}(2022)Zhang, Udrea, and Wang]{2-RN9967}
Zhang,~Y.; Udrea,~F.; Wang,~H. Multidimensional device architectures for
  efficient power electronics. \emph{Nature Electronics} \textbf{2022},
  \emph{5}, 723--734\relax
\mciteBstWouldAddEndPuncttrue
\mciteSetBstMidEndSepPunct{\mcitedefaultmidpunct}
{\mcitedefaultendpunct}{\mcitedefaultseppunct}\relax
\EndOfBibitem
\bibitem[Vandersande(1976)]{3-RN9942}
Vandersande,~J.~W. Boundary scattering in five natural type-II a diamonds.
  \emph{Physical Review B} \textbf{1976}, \emph{13}, 4560--4567\relax
\mciteBstWouldAddEndPuncttrue
\mciteSetBstMidEndSepPunct{\mcitedefaultmidpunct}
{\mcitedefaultendpunct}{\mcitedefaultseppunct}\relax
\EndOfBibitem
\bibitem[Wei \latin{et~al.}(1993)Wei, Kuo, Thomas, Anthony, and
  Banholzer]{4-RN9969}
Wei,~L.; Kuo,~P.~K.; Thomas,~R.~L.; Anthony,~T.~R.; Banholzer,~W.~F. Thermal
  conductivity of isotopically modified single crystal diamond. \emph{Physical
  Review Letters} \textbf{1993}, \emph{70}, 3764--3767\relax
\mciteBstWouldAddEndPuncttrue
\mciteSetBstMidEndSepPunct{\mcitedefaultmidpunct}
{\mcitedefaultendpunct}{\mcitedefaultseppunct}\relax
\EndOfBibitem
\bibitem[Onn \latin{et~al.}(1992)Onn, Witek, Qiu, Anthony, and
  Banholzer]{5-RN9968}
Onn,~D.~G.; Witek,~A.; Qiu,~Y.~Z.; Anthony,~T.~R.; Banholzer,~W.~F. Some
  aspects of the thermal conductivity of isotopically enriched diamond single
  crystals. \emph{Physical Review Letters} \textbf{1992}, \emph{68},
  2806--2809\relax
\mciteBstWouldAddEndPuncttrue
\mciteSetBstMidEndSepPunct{\mcitedefaultmidpunct}
{\mcitedefaultendpunct}{\mcitedefaultseppunct}\relax
\EndOfBibitem
\bibitem[Inyushkin \latin{et~al.}(2018)Inyushkin, Taldenkov, Ralchenko,
  Bolshakov, Koliadin, and Katrusha]{6-RN9948}
Inyushkin,~A.~V.; Taldenkov,~A.~N.; Ralchenko,~V.~G.; Bolshakov,~A.~P.;
  Koliadin,~A.~V.; Katrusha,~A.~N. Thermal conductivity of high purity
  synthetic single crystal diamonds. \emph{Physical Review B} \textbf{2018},
  \emph{97}, 144305\relax
\mciteBstWouldAddEndPuncttrue
\mciteSetBstMidEndSepPunct{\mcitedefaultmidpunct}
{\mcitedefaultendpunct}{\mcitedefaultseppunct}\relax
\EndOfBibitem
\bibitem[Kucsko \latin{et~al.}(2013)Kucsko, Maurer, Yao, Kubo, Noh, Lo, Park,
  and Lukin]{7-RN9949}
Kucsko,~G.; Maurer,~P.~C.; Yao,~N.~Y.; Kubo,~M.; Noh,~H.~J.; Lo,~P.~K.;
  Park,~H.; Lukin,~M.~D. Nanometre-scale thermometry in a living cell.
  \emph{Nature} \textbf{2013}, \emph{500}, 54--58\relax
\mciteBstWouldAddEndPuncttrue
\mciteSetBstMidEndSepPunct{\mcitedefaultmidpunct}
{\mcitedefaultendpunct}{\mcitedefaultseppunct}\relax
\EndOfBibitem
\bibitem[Taylor \latin{et~al.}(2008)Taylor, Cappellaro, Childress, Jiang,
  Budker, Hemmer, Yacoby, Walsworth, and Lukin]{8-RN9950}
Taylor,~J.~M.; Cappellaro,~P.; Childress,~L.; Jiang,~L.; Budker,~D.;
  Hemmer,~P.~R.; Yacoby,~A.; Walsworth,~R.; Lukin,~M.~D. High-sensitivity
  diamond magnetometer with nanoscale resolution. \emph{Nature Physics}
  \textbf{2008}, \emph{4}, 810--816\relax
\mciteBstWouldAddEndPuncttrue
\mciteSetBstMidEndSepPunct{\mcitedefaultmidpunct}
{\mcitedefaultendpunct}{\mcitedefaultseppunct}\relax
\EndOfBibitem
\bibitem[Balasubramanian \latin{et~al.}(2008)Balasubramanian, Chan, Kolesov,
  Al-Hmoud, Tisler, Shin, Kim, Wojcik, Hemmer, Krueger, Hanke, Leitenstorfer,
  Bratschitsch, Jelezko, and Wrachtrup]{9-RN9951}
Balasubramanian,~G.; Chan,~I.~Y.; Kolesov,~R.; Al-Hmoud,~M.; Tisler,~J.;
  Shin,~C.; Kim,~C.; Wojcik,~A.; Hemmer,~P.~R.; Krueger,~A.; Hanke,~T.;
  Leitenstorfer,~A.; Bratschitsch,~R.; Jelezko,~F.; Wrachtrup,~J. Nanoscale
  imaging magnetometry with diamond spins under ambient conditions.
  \emph{Nature} \textbf{2008}, \emph{455}, 648--651\relax
\mciteBstWouldAddEndPuncttrue
\mciteSetBstMidEndSepPunct{\mcitedefaultmidpunct}
{\mcitedefaultendpunct}{\mcitedefaultseppunct}\relax
\EndOfBibitem
\bibitem[Mochalin \latin{et~al.}(2012)Mochalin, Shenderova, Ho, and
  Gogotsi]{10-RN9898}
Mochalin,~V.~N.; Shenderova,~O.; Ho,~D.; Gogotsi,~Y. The properties and
  applications of nanodiamonds. \emph{Nature Nanotechnology} \textbf{2012},
  \emph{7}, 11--23\relax
\mciteBstWouldAddEndPuncttrue
\mciteSetBstMidEndSepPunct{\mcitedefaultmidpunct}
{\mcitedefaultendpunct}{\mcitedefaultseppunct}\relax
\EndOfBibitem
\bibitem[Le~Sage \latin{et~al.}(2013)Le~Sage, Arai, Glenn, DeVience, Pham,
  Rahn-Lee, Lukin, Yacoby, Komeili, and Walsworth]{12-RN6403}
Le~Sage,~D.; Arai,~K.; Glenn,~D.~R.; DeVience,~S.~J.; Pham,~L.~M.;
  Rahn-Lee,~L.; Lukin,~M.~D.; Yacoby,~A.; Komeili,~A.; Walsworth,~R.~L. Optical
  magnetic imaging of living cells. \emph{Nature} \textbf{2013}, \emph{496},
  486--489\relax
\mciteBstWouldAddEndPuncttrue
\mciteSetBstMidEndSepPunct{\mcitedefaultmidpunct}
{\mcitedefaultendpunct}{\mcitedefaultseppunct}\relax
\EndOfBibitem
\bibitem[Schmitt \latin{et~al.}(2017)Schmitt, Gefen, Stuerner, Unden, Wolff,
  Mueller, Scheuer, Naydenov, Markham, and Pezzagna]{11-RN8010}
Schmitt,~S.; Gefen,~T.; Stuerner,~F.~M.; Unden,~T.; Wolff,~G.; Mueller,~C.;
  Scheuer,~J.; Naydenov,~B.; Markham,~M.; Pezzagna,~S. Submillihertz magnetic
  spectroscopy performed with a nanoscale quantum sensor. \emph{Science}
  \textbf{2017}, \emph{356}, 832--837\relax
\mciteBstWouldAddEndPuncttrue
\mciteSetBstMidEndSepPunct{\mcitedefaultmidpunct}
{\mcitedefaultendpunct}{\mcitedefaultseppunct}\relax
\EndOfBibitem
\bibitem[Sipahigil \latin{et~al.}(2016)Sipahigil, Evans, Sukachev, Burek,
  Borregaard, Bhaskar, Nguyen, Pacheco, Atikian, Meuwly, Camacho, Jelezko,
  Bielejec, Park, Lončar, and Lukin]{13-RN8651}
Sipahigil,~A. \latin{et~al.}  An integrated diamond nanophotonics platform for
  quantum-optical networks. \emph{Science} \textbf{2016}, \emph{354},
  847--850\relax
\mciteBstWouldAddEndPuncttrue
\mciteSetBstMidEndSepPunct{\mcitedefaultmidpunct}
{\mcitedefaultendpunct}{\mcitedefaultseppunct}\relax
\EndOfBibitem
\bibitem[Häußler \latin{et~al.}(2019)Häußler, Benedikter, Bray, Regan,
  Dietrich, Twamley, Aharonovich, Hunger, and Kubanek]{14-RN5152}
Häußler,~S.; Benedikter,~J.; Bray,~K.; Regan,~B.; Dietrich,~A.; Twamley,~J.;
  Aharonovich,~I.; Hunger,~D.; Kubanek,~A. Diamond photonics platform based on
  silicon vacancy centers in a single-crystal diamond membrane and a fiber
  cavity. \emph{Physical Review B} \textbf{2019}, \emph{99}, 165310\relax
\mciteBstWouldAddEndPuncttrue
\mciteSetBstMidEndSepPunct{\mcitedefaultmidpunct}
{\mcitedefaultendpunct}{\mcitedefaultseppunct}\relax
\EndOfBibitem
\bibitem[Dolde \latin{et~al.}(2011)Dolde, Fedder, Doherty, Nöbauer, Rempp,
  Balasubramanian, Wolf, Reinhard, Hollenberg, Jelezko, and
  Wrachtrup]{15-RN4908}
Dolde,~F.; Fedder,~H.; Doherty,~M.~W.; Nöbauer,~T.; Rempp,~F.;
  Balasubramanian,~G.; Wolf,~T.; Reinhard,~F.; Hollenberg,~L. C.~L.;
  Jelezko,~F.; Wrachtrup,~J. Electric-field sensing using single diamond spins.
  \emph{Nature Physics} \textbf{2011}, \emph{7}, 459--463\relax
\mciteBstWouldAddEndPuncttrue
\mciteSetBstMidEndSepPunct{\mcitedefaultmidpunct}
{\mcitedefaultendpunct}{\mcitedefaultseppunct}\relax
\EndOfBibitem
\bibitem[Acosta \latin{et~al.}(2010)Acosta, Bauch, Ledbetter, Waxman, Bouchard,
  and Budker]{16-RN9953}
Acosta,~V.~M.; Bauch,~E.; Ledbetter,~M.~P.; Waxman,~A.; Bouchard,~L.~S.;
  Budker,~D. Temperature Dependence of the Nitrogen-Vacancy Magnetic Resonance
  in Diamond. \emph{Physical Review Letters} \textbf{2010}, \emph{104},
  070801\relax
\mciteBstWouldAddEndPuncttrue
\mciteSetBstMidEndSepPunct{\mcitedefaultmidpunct}
{\mcitedefaultendpunct}{\mcitedefaultseppunct}\relax
\EndOfBibitem
\bibitem[Häußler \latin{et~al.}(2017)Häußler, Thiering, Dietrich, Waasem,
  Teraji, Isoya, Iwasaki, Hatano, Jelezko, Gali, and Kubanek]{17-RN6280}
Häußler,~S.; Thiering,~G.; Dietrich,~A.; Waasem,~N.; Teraji,~T.; Isoya,~J.;
  Iwasaki,~T.; Hatano,~M.; Jelezko,~F.; Gali,~A.; Kubanek,~A. Photoluminescence
  excitation spectroscopy of SiV- and GeV- color center in diamond. \emph{New
  Journal of Physics} \textbf{2017}, \emph{19}, 063036\relax
\mciteBstWouldAddEndPuncttrue
\mciteSetBstMidEndSepPunct{\mcitedefaultmidpunct}
{\mcitedefaultendpunct}{\mcitedefaultseppunct}\relax
\EndOfBibitem
\bibitem[Fan \latin{et~al.}(2018)Fan, Cojocaru, Becker, Fedotov, Alkahtani,
  Alajlan, Blakley, Rezaee, Lyamkina, Palyanov, Borzdov, Yang, Zheltikov,
  Hemmer, and Akimov]{18-RN6043}
Fan,~J.-W.; Cojocaru,~I.; Becker,~J.; Fedotov,~I.~V.; Alkahtani,~M. H.~A.;
  Alajlan,~A.; Blakley,~S.; Rezaee,~M.; Lyamkina,~A.; Palyanov,~Y.~N.;
  Borzdov,~Y.~M.; Yang,~Y.-P.; Zheltikov,~A.; Hemmer,~P.; Akimov,~A.~V.
  Germanium-Vacancy Color Center in Diamond as a Temperature Sensor. \emph{ACS
  Photonics} \textbf{2018}, \emph{5}, 765--770\relax
\mciteBstWouldAddEndPuncttrue
\mciteSetBstMidEndSepPunct{\mcitedefaultmidpunct}
{\mcitedefaultendpunct}{\mcitedefaultseppunct}\relax
\EndOfBibitem
\bibitem[Clark \latin{et~al.}(1995)Clark, Kanda, Kiflawi, and
  Sittas]{19-RN6153}
Clark,~C.~D.; Kanda,~H.; Kiflawi,~I.~I.; Sittas,~G. Silicon defects in diamond.
  \emph{Phys Rev B Condens Matter} \textbf{1995}, \emph{51}, 16681--16688\relax
\mciteBstWouldAddEndPuncttrue
\mciteSetBstMidEndSepPunct{\mcitedefaultmidpunct}
{\mcitedefaultendpunct}{\mcitedefaultseppunct}\relax
\EndOfBibitem
\bibitem[Hepp \latin{et~al.}(2014)Hepp, Müller, Waselowski, Becker, Pingault,
  Sternschulte, Steinmüller-Nethl, Gali, Maze, Atatüre, and
  Becher]{20-RN7795}
Hepp,~C.; Müller,~T.; Waselowski,~V.; Becker,~J.; Pingault,~B.;
  Sternschulte,~H.; Steinmüller-Nethl,~D.; Gali,~A.; Maze,~J.; Atatüre,~M.;
  Becher,~C. Electronic Structure of the Silicon Vacancy Color Center in
  Diamond. \emph{Physical review letters} \textbf{2014}, \emph{112},
  036405\relax
\mciteBstWouldAddEndPuncttrue
\mciteSetBstMidEndSepPunct{\mcitedefaultmidpunct}
{\mcitedefaultendpunct}{\mcitedefaultseppunct}\relax
\EndOfBibitem
\bibitem[Meesala \latin{et~al.}(2018)Meesala, Sohn, Pingault, Shao, Atikian,
  Holzgrafe, Gündoğan, Stavrakas, Sipahigil, Chia, Evans, Burek, Zhang, Wu,
  Pacheco, Abraham, Bielejec, Lukin, Atatüre, and Lončar]{21-RN4053}
Meesala,~S. \latin{et~al.}  Strain engineering of the silicon-vacancy center in
  diamond. \emph{Physical Review B} \textbf{2018}, \emph{97}, 205444\relax
\mciteBstWouldAddEndPuncttrue
\mciteSetBstMidEndSepPunct{\mcitedefaultmidpunct}
{\mcitedefaultendpunct}{\mcitedefaultseppunct}\relax
\EndOfBibitem
\bibitem[Lin \latin{et~al.}(2018)Lin, Lu, Yang, Tian, Gao, Sun, Dong, Zhong,
  Hu, and Shan]{22-RN4418}
Lin,~C.-N.; Lu,~Y.-J.; Yang,~X.; Tian,~Y.-Z.; Gao,~C.-J.; Sun,~J.-L.; Dong,~L.;
  Zhong,~F.; Hu,~W.-D.; Shan,~C.-X. Diamond-Based All-Carbon Photodetectors for
  Solar-Blind Imaging. \emph{Advanced Optical Materials} \textbf{2018},
  \emph{6}, 1800068\relax
\mciteBstWouldAddEndPuncttrue
\mciteSetBstMidEndSepPunct{\mcitedefaultmidpunct}
{\mcitedefaultendpunct}{\mcitedefaultseppunct}\relax
\EndOfBibitem
\bibitem[Sternschulte \latin{et~al.}(1995)Sternschulte, Thonke, Gerster,
  Limmer, Sauer, Spitzer, and Münzinger]{23-RN4878}
Sternschulte,~H.; Thonke,~K.; Gerster,~J.; Limmer,~W.; Sauer,~R.; Spitzer,~J.;
  Münzinger,~P.~C. Uniaxial stress and Zeeman splitting of the 1.681 eV
  optical center in a homoepitaxial CVD diamond film. \emph{Diamond and Related
  Materials} \textbf{1995}, \emph{4}, 1189--1192\relax
\mciteBstWouldAddEndPuncttrue
\mciteSetBstMidEndSepPunct{\mcitedefaultmidpunct}
{\mcitedefaultendpunct}{\mcitedefaultseppunct}\relax
\EndOfBibitem
\bibitem[Sternschulte \latin{et~al.}(1994)Sternschulte, Thonke, Sauer,
  Münzinger, and Michler]{24-RN6127}
Sternschulte,~H.; Thonke,~K.; Sauer,~R.; Münzinger,~P.~C.; Michler,~P.
  1.681-eV luminescence center in chemical-vapor-deposited homoepitaxial
  diamond films. \emph{Physical Review B} \textbf{1994}, \emph{50},
  14554--14560\relax
\mciteBstWouldAddEndPuncttrue
\mciteSetBstMidEndSepPunct{\mcitedefaultmidpunct}
{\mcitedefaultendpunct}{\mcitedefaultseppunct}\relax
\EndOfBibitem
\bibitem[Liu and Kauffman(1999)Liu, and Kauffman]{25-RN9899}
Liu,~C.~S.; Kauffman,~J.~F. Photoluminescence and interfacial heat transfer in
  gallium arsenide. \emph{Applied Physics Letters} \textbf{1999}, \emph{75},
  1434--1436\relax
\mciteBstWouldAddEndPuncttrue
\mciteSetBstMidEndSepPunct{\mcitedefaultmidpunct}
{\mcitedefaultendpunct}{\mcitedefaultseppunct}\relax
\EndOfBibitem
\bibitem[Dobal \latin{et~al.}(1994)Dobal, Bist, Mehta, and Jain]{26-RN6332}
Dobal,~P.~S.; Bist,~H.~D.; Mehta,~S.~K.; Jain,~R.~K. Laser heating and
  photoluminescence in $GaAs$ and $Al_{x}Ga_{1-x}As$. \emph{Applied Physics
  Letters} \textbf{1994}, \emph{65}, 2469--2471\relax
\mciteBstWouldAddEndPuncttrue
\mciteSetBstMidEndSepPunct{\mcitedefaultmidpunct}
{\mcitedefaultendpunct}{\mcitedefaultseppunct}\relax
\EndOfBibitem
\bibitem[Varshni(1967)]{27-RN9957}
Varshni,~Y.~P. Temperature dependence of the energy gap in semiconductors.
  \emph{Physica} \textbf{1967}, \emph{34}, 149--154\relax
\mciteBstWouldAddEndPuncttrue
\mciteSetBstMidEndSepPunct{\mcitedefaultmidpunct}
{\mcitedefaultendpunct}{\mcitedefaultseppunct}\relax
\EndOfBibitem
\bibitem[Bergman \latin{et~al.}(2004)Bergman, Chen, Morrison, Huso, and
  Purdy]{28-RN7944}
Bergman,~L.; Chen,~X.~B.; Morrison,~J.~L.; Huso,~J.; Purdy,~A.~P.
  Photoluminescence dynamics in ensembles of wide-band-gap nanocrystallites and
  powders. \emph{Journal of Applied Physics} \textbf{2004}, \emph{96},
  675--682\relax
\mciteBstWouldAddEndPuncttrue
\mciteSetBstMidEndSepPunct{\mcitedefaultmidpunct}
{\mcitedefaultendpunct}{\mcitedefaultseppunct}\relax
\EndOfBibitem
\bibitem[Yang \latin{et~al.}(2006)Yang, Yan, Fu, Yang, Xia, Xu, Zuo, and
  Li]{29-RN3582}
Yang,~Y.; Yan,~H.; Fu,~Z.; Yang,~B.; Xia,~L.; Xu,~Y.; Zuo,~J.; Li,~F.
  Photoluminescence investigation based on laser heating effect in ZnO-ordered
  nanostructures. \emph{Journal of Physical Chemistry B} \textbf{2006},
  \emph{110}, 846--852\relax
\mciteBstWouldAddEndPuncttrue
\mciteSetBstMidEndSepPunct{\mcitedefaultmidpunct}
{\mcitedefaultendpunct}{\mcitedefaultseppunct}\relax
\EndOfBibitem
\bibitem[Li \latin{et~al.}(1997)Li, Huang, Malikova, and Pollak]{30-RN9900}
Li,~C.~F.; Huang,~Y.~S.; Malikova,~L.; Pollak,~F.~H. Temperature dependence of
  the energies and broadening parameters of the interband excitonic transitions
  in wurtzite GaN. \emph{Physical Review B} \textbf{1997}, \emph{55},
  9251--9254\relax
\mciteBstWouldAddEndPuncttrue
\mciteSetBstMidEndSepPunct{\mcitedefaultmidpunct}
{\mcitedefaultendpunct}{\mcitedefaultseppunct}\relax
\EndOfBibitem
\bibitem[Feng and Schwartz(1993)Feng, and Schwartz]{31-RN7709}
Feng,~T.; Schwartz,~B. Characteristics and origin of the 1.681 eV luminescence
  center in chemical‐vapor‐deposited diamond films. \emph{Journal of
  Applied Physics} \textbf{1993}, \emph{73}, 1415--1425\relax
\mciteBstWouldAddEndPuncttrue
\mciteSetBstMidEndSepPunct{\mcitedefaultmidpunct}
{\mcitedefaultendpunct}{\mcitedefaultseppunct}\relax
\EndOfBibitem
\bibitem[Neu \latin{et~al.}(2013)Neu, Hepp, Hauschild, Gsell, Fischer,
  Sternschulte, Steinmüller-Nethl, Schreck, and Becher]{32-RN8547}
Neu,~E.; Hepp,~C.; Hauschild,~M.; Gsell,~S.; Fischer,~M.; Sternschulte,~H.;
  Steinmüller-Nethl,~D.; Schreck,~M.; Becher,~C. Low-temperature
  investigations of single silicon vacancy colour centres in diamond. \emph{New
  Journal of Physics} \textbf{2013}, \emph{15}, 043005\relax
\mciteBstWouldAddEndPuncttrue
\mciteSetBstMidEndSepPunct{\mcitedefaultmidpunct}
{\mcitedefaultendpunct}{\mcitedefaultseppunct}\relax
\EndOfBibitem
\bibitem[Morelli \latin{et~al.}(1993)Morelli, Perry, and Farmer]{33-RN9959}
Morelli,~D.~T.; Perry,~T.~A.; Farmer,~J.~W. Phonon scattering in lightly
  neutron-irradiated diamond. \emph{Physical Review B} \textbf{1993},
  \emph{47}, 131--139\relax
\mciteBstWouldAddEndPuncttrue
\mciteSetBstMidEndSepPunct{\mcitedefaultmidpunct}
{\mcitedefaultendpunct}{\mcitedefaultseppunct}\relax
\EndOfBibitem
\bibitem[Gao \latin{et~al.}(2022)Gao, Lai, Sun, Liu, Lin, Tan, Shan, and
  Zhang]{34-RN9965}
Gao,~Y.-F.; Lai,~J.-M.; Sun,~Y.-J.; Liu,~X.-L.; Lin,~C.-N.; Tan,~P.-H.;
  Shan,~C.-X.; Zhang,~J. Charge State Manipulation of NV Centers in Diamond
  under Phonon-Assisted Anti-Stokes Excitation of $NV^0$. \emph{ACS Photonics}
  \textbf{2022}, \emph{9}, 1605--1613\relax
\mciteBstWouldAddEndPuncttrue
\mciteSetBstMidEndSepPunct{\mcitedefaultmidpunct}
{\mcitedefaultendpunct}{\mcitedefaultseppunct}\relax
\EndOfBibitem
\end{mcitethebibliography}
\end{document}